\documentclass[useAMS,usenatbib]{mn2e}

\usepackage{epsfig}
\usepackage{longtable}
\usepackage{times} 
\newcommand{\sw}{$Swift$}

\def \src {\mbox{IGR~J16493--4348}}
\def \igr17544 {\mbox{IGR~J17544$-$2619}}
\def \xte17391 {\mbox{XTE~J1739$-$302}}
\def \IGRJ16479 {\mbox{IGR~J16479$-$4514}}
\def \my11215 {\mbox{IGR~J11215$-$5952}}

\def \sw {{\em Swift}}

\def \ferg {erg cm$^{-2}$ s$^{-1}$}
\def \hcm {\hbox {\ifmmode $ atom cm$^{-2}\else atom cm$^{-2}$\fi}}

\def \apj {ApJ}

\def \aap {A\&A}

\title[Orbital period in IGR~J16493-4348]
{The \emph{Swift}-BAT hard X-ray sky monitoring unveils the orbital period
 of the HMXB  IGR J16493--4348}

\author[G.\ Cusumano et al.]{G.\ Cusumano$^{1}$, V.\ La Parola$^{1}$, 
P.\ Romano$^{1}$, A.\ Segreto $^{1}$, S.\ Vercellone$^{1}$, G. Chincarini$^{2,3}$ \\
$^{1}$INAF, Istituto di Astrofisica Spaziale e Fisica Cosmica,
        Via U.\ La Malfa 153, I-90146 Palermo, Italy\\
$^{2}$ INAF-Osservatorio Astronomico di Brera, I-23807 Merate (LC), Italy\\
$^{3}$ Universit\'a degli Studi di Milano, Bicocca, I-20126 Milano, Italy\\
}

\begin{document}

\date{}

\pagerange{\pageref{firstpage}--\pageref{lastpage}} \pubyear{2010}

\maketitle

\label{firstpage}

\begin{abstract}
\src\ is a supergiant high mass X-ray binary discovered by INTEGRAL in 2004.
The source is detected at a significance level of $\sim21$
standard deviations in the \sw-BAT survey data collected during the first 54 
months of the \sw\ mission.
The timing analysis reveals an orbital 
period of $\sim$6.78 days and the  presence of a full eclipse of the compact object. 
The dynamical range (variability up to a factor $\sim$20) observed 
during the BAT monitoring suggests that \src\ is a wind-fed system. The derived 
semi-major axis of the binary system is $\sim55 R_{\sun}$ with an orbit eccentricity
lower than 0.15.

\end{abstract}

\begin{keywords}
X-rays: binaries -- X-rays: individual: IGR~J16493$-$4348. 

\noindent
Facility: {\it Swift}

\end{keywords}


        \section{Introduction\label{sfxt7:intro}}

The IBIS/ISGRI telescope \citep{Ubertini03,Lebrun03} on board 
the INTEGRAL satellite \citep{Winkler03} has detected a large number of new 
supergiant high mass X-ray binaries (sgHMXB) characterized by a high 
absorption column (N$_{\rm H}> 10^{23}$ cm$^2$) and/or by fast
bright transient events. This result was achieved thanks to a continuous  
monitoring of the Galactic plane with deep exposure of the Galactic centre
and thanks to a combination of ISGRI large field of view, good sensitivity,
and wide energy range. 
The fourth IBIS/ISGRI survey catalogue \citep{Bird10} reports the discovery of 
32 new objects that have been classified as HMXB and  84 unidentified sources 
with a Galactic latitude $|b| < 5^{\circ}$.

The Burst Alert Telescope (BAT, \citealp{Barthelmy2005:BAT})
on board \sw\ \citep{Gehrels2004mn} has been performing a 
continuous monitoring of the sky in the hard X-ray energy range (15--150 keV) since 
November 2004.
The telescope,  thanks to its large field of view 
(1.4 steradian half coded) and its pointing strategy, covers a
fraction between 50\% and 80\% of the sky every day.
This has allowed the detection of many of the new INTEGRAL
HMXBs \citep[e.g.][]{Cusumano2010} and the collection of their long
term light curves and of their spectral energy distributions. 
The long and continuous monitoring of these sources allows to 
investigate the intrinsic emission variability, 
to search for long periodicities (orbital periods) and to discover the presence 
of eclipse events. The role of \sw-BAT is therefore fundamental to unveil 
the nature and the geometry of these binary systems. 

In this Letter we analyze the soft and hard X-ray data collected by \sw\ on 
IGR~J16493--4348. This source was discovered by INTEGRAL in 2004 
\citep{Bird2004} and it was initially associated with the radio pulsar 
PSR~J1649--4349 because of a spatial coincidence. A later INTEGRAL observation 
with a deep exposure allowed to reduce the positional uncertainty and to reject 
the pulsar association \citep{Grebenev2005}.
A follow-up observation with {\em Chandra} found a soft X-ray counterpart at 
RA(J2000) = 16$^{h}$ 49$^{m}$ 26.92$^{s}$; Dec(J2000) = -43$^{\circ}$ 49' 8.96'' 
\citep{Kuiper2005} allowing 
the optical association with 2MASS~J1642695--4349090, a B0.5 Ib
supergiant \citep{Nespoli2008}. The {\em Chandra} observation also revealed strong 
evidence for variability.
The IGR~J16493--4348 energy distribution extracted from two  RXTE observations 
\citep{markwardt05} was modeled with a highly absorbed power law 
(N$_{\rm H}\sim10^{23}$ cm$^{-2}$)  with a photon 
index of $\sim$1.4 and a 20--40 keV flux of 2.1$ \times 10^{-11}$ \ferg.
A spectral analysis using non simultaneous data from \sw-XRT and INTEGRAL \citep{hill07}
showed the presence of a cutoff at $\sim15$ keV.
The information derived from the X-ray observations and the identification of 
the spectral type of the optical counterpart allowed the classification of 
\src\ as a supergiant High Mass X-ray Binary (sgHMXB).
 
This Letter is organized as follows. Section 2 describes the data reduction.
 Section 3 reports on the timing analysis and
in Sect. 4 we discuss our results. 

        \section{Data Reduction\label{sfxt7:data}}

\begin{figure}
\begin{center}
\centerline{\includegraphics[width=7.5cm,angle=0]{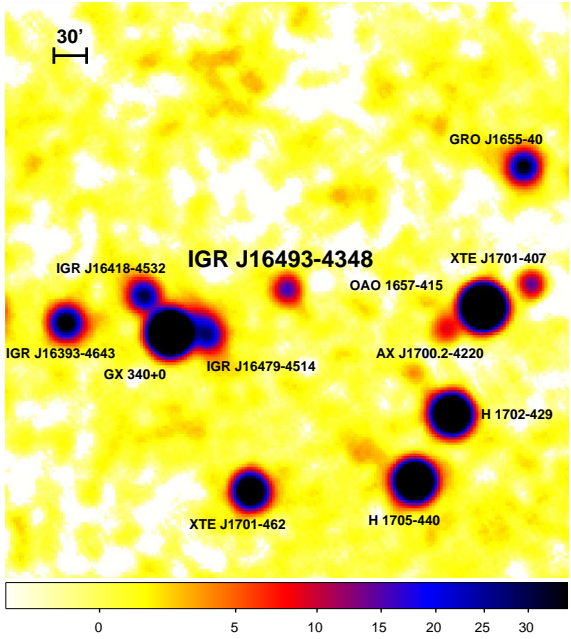}}
\caption[IGRJ16493-4348 sky map]{{\bf Top panel}: 15--50 keV significance map in the 
neighborhood of IGR~J16493$-$4348. The color-bar represents the significance levels.
                }
                \label{map} 
        \end{center}
        \end{figure}

The raw BAT survey data of the first 54 months of the \sw\ 
mission were retrieved from the HEASARC public archive\footnote{http://heasarc.gsfc.nasa.gov/docs/archive.html} and
processed with a dedicated software \citep{segreto10}, that performs screening, 
mosaicking and source detection on BAT data and produces spectra and light curves for any given 
sky position. Figure~\ref{map} shows the 15--50 keV significance sky map (exposure time of 17.7 Ms,
duty cycle $\sim$ 13\%) centered in
the direction of IGR~J16493$-$4348. The source is clearly detected at a significance 
of 20.8 standard deviations.
 The light curve of \src\, was extracted in the 15--50 keV energy range with the 
maximum available time resolution 
($\sim 300$ s), subtracting the contamination of the nearby
sources as detailed in  \citet{segreto10}.  The time tag of each bin, defined as the central time of the bin interval,
 was corrected to the Solar system barycentre (SSB) by 
using the task {\sc EARTH2SUN}. 
 
We also re-analyzed the data of the \sw-XRT \citep{Burrows2005:XRTmn} IGR J16493--4348 observation
performed on 2006 March 11 (ObsId 00030379002), for a total exposure time of 5.6 ks. 
The data were processed with
standard procedures ({\sc xrtpipeline} v.0.12.4), filtering and screening criteria, using {\sc ftools} in the {\sc heasoft}
package (v 6.8). The source was observed in Photon Counting mode \citep{hill04}
that provides a time resolution of 2.5 s.

 We adopted standard grade filtering
0--12. The source events were extracted from a circular region of 20 pixels radius 
(1 pixel=2.36'')
centered on the source position as determined with {\sc xrtcentroid}. The event arrival times were 
converted to the SSB with the task {\sc barycorr}. 

\begin{figure}[t]
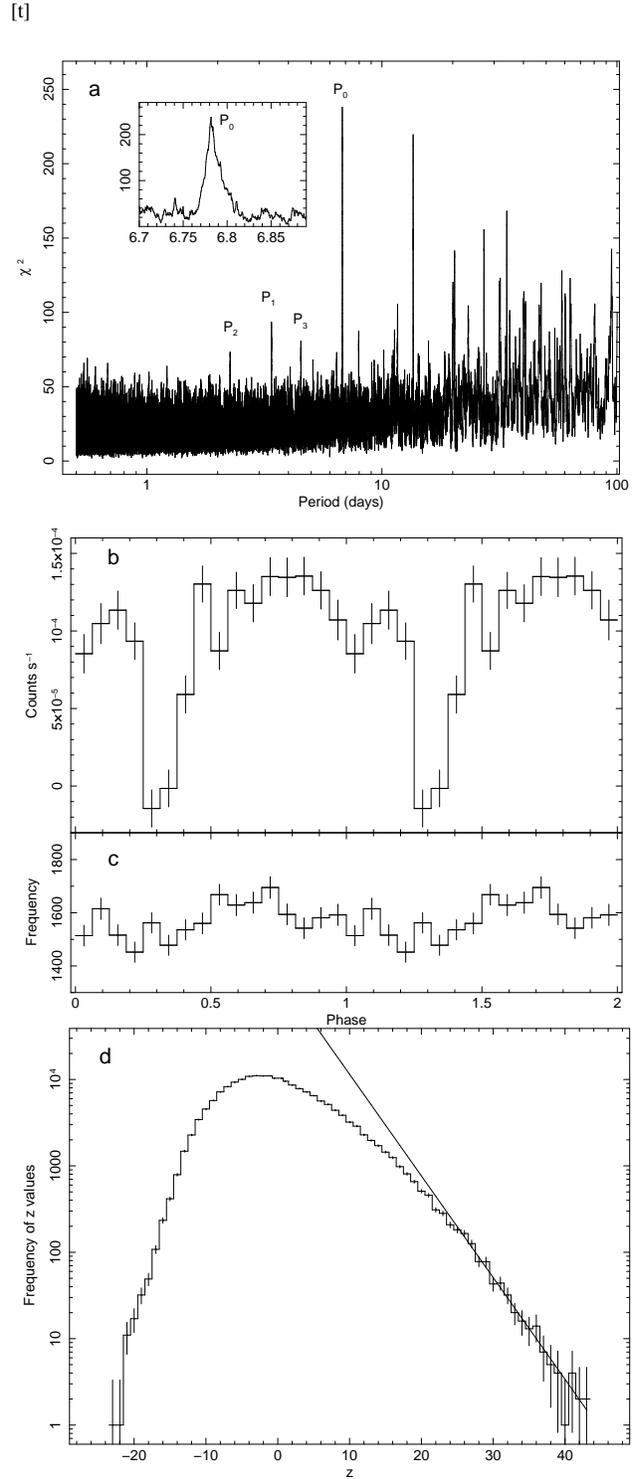

\begin{center}
\centerline{\includegraphics[width=6cm,angle=270]{figure2a.ps}}
\centerline{\includegraphics[width=6.8cm,angle=270]{figure2b.ps}}
\centerline{\includegraphics[width=6cm,angle=270]{figure2c.ps}}

\caption[]{{\bf a}: Periodogram of \sw-BAT (15--50\,keV) data for 
IGR~J16493--4348. P$_0$, P$_1$, P$_2$ and P$_3$ are defined in Section 3. The inset shows a close-up view of the $\chi^2$ 
distribution around P$_0$.
{\bf b}: Light curve folded at a period P$_0= 6.783$\,days, with 16 phase 
bins.
{\bf c}: Phase coverage of the light curve bins. The errors are the square root of the number of bins in each phase interval 
{\bf d}: Distribution of the z=$\chi^2$-F$_{\chi}$(P) values extracted in the period range between 0.5 and 10 days excluding the 
z values obtained at P$_0$, P$_1$, P$_2$ and P$_3$.
The continuous line is the best fit obtained with an
exponential model  applied to the tail of the distribution (z$>$25).                }               
		\label{period} 
        \end{center}
        \end{figure}

        \section{Timing analysis\label{sfxt7:timing}}
\subsection{BAT survey data}
The average count rate in the BAT light curve is $1.03\times 10^{-4}$ count s$^{-1}$.
When considering the light curve at the highest resolution the maximum
of the deviation from the average rate is about $7\sigma$, corresponding to an increase
in rate of a factor $\sim20$.

We analyzed the long term BAT light curve to search for intensity modulations
by applying a folding technique and
searching in the 0.5--100\,d period range.
The period resolution \citep{buccheri85} is given by P$^{2}/(N \,\Delta T)$,
where $N=16$ is the number of trial profile phase bins, and $\Delta T=$140,213,639.0  s is the 
data time span. 
The average rate in each phase bin was evaluated by weighting the light curve rates 
by the inverse square of the corresponding statistical error
\begin{equation}
R_j=\frac{\sum{r_i/er_i^2}}{\sum{1/er_i^2}}  
\end{equation}
where $R_j$ is the average rate in the j-th phase bin of the trial profile, $r_i$ are the
rate of the light curve whose phase fall  into the j-th phase bin and $er_i$ are the
corresponding statistical errors. The error on $R_j$ is
$(\sqrt{\sum{1/er_i^2}})^{-1}$. The weighting procedure was adopted to deal with the large span 
of $er_i$ and it is justified because the BAT data are background dominated. 

Figure~\ref{period} (a) shows the periodogram where several features emerge. 
The highest feature, with a $\chi^2$ value of $\sim240$, is P$_0=6.782\pm0.005$ d,
where the error is evaluated as the standard deviation of the gaussian that best fits 
the $\chi^2$ values around the P$_0$ peak with $\chi^2 > 150$.
We also see other evident features at periods multiple of P$_0$ (2P$_0$, 3P$_0$, 4P$_0$, 
5P$_0$ in Fig.~\ref{period}). The pulsed profile (Fig.~\ref{period}, b) 
folded at P$_0$ with T$_{\rm epoch}$=54173.757 MJD, shows a flat intensity level, abruptly 
broken by a deep full eclipse.  The phase coverage
of the light curve bins (Fig.~\ref{period}, c) shows that this dip is not due to an accidental 
under-sampling of the light curve at these phases.
The centroid of the eclipse, evaluated by fitting the 
data around the dip with a Gaussian model, is at phase $0.319\pm 0.015$ corresponding to 
T$_{\rm eclipse}=(54175.92\pm 0.10)\pm n \times $P$_0$ MJD.

As a consequence of the time variability of the source and of the presence of a periodic 
signal, the average $\chi^2$ in 
the periodogram is far from the average value expected for white noise $(N-1)$ and 
the $\chi^2$ statistics cannot  be applied to evaluate the significance of the detected 
periodicity. 
Therefore, we applied two alternative methods to evaluate the significance of P$_0$. 

\begin{enumerate}
\item  We fit the periodogram with a second order polynomial [F$_{\chi}$(P)] and subtracted the trend from 
the $\chi^2$ distribution. The value of z=$\chi^2$-F$_{\chi}$(P) at P$_0$ is 196.6. We therefore built 
the histogram of the z distribution (Fig.~\ref{period}, d) 
from 0.5 to 10 d excluding the interval  around P$_0$ and those around P$_1$, P$_2$, and P$_3$ as
marked in Fig.~\ref{period} (a).
The latter features are not noise fluctuations, but are tied to P$_0$: P$_1$ (=P$_0/2$) is
due to the presence of the deep eclipse, still visible in the light curve folded at half of the
period; P$_2$ ($\sim 2.27$d$=(1/$P$_0 + 1/$P$_1)^{-1}$)  and P$_3$ ($\sim 4.54$d$=[(1/$P$_0 + 1/$P$_1)/2]^{-1}$) are 
due to beat frequencies between P$_0$ and P$_1$.
We fit the resulting distribution for z$>25$  with an exponential 
function and evaluated the integral of the best-fit function beyond z=196.6.
This integral yields a number of chance occurrences due to noise of 
$4.6\times10^{-18}$, corresponding to a significance for the detected feature of 
$\sim8.5$ standard deviations in Gaussian statistics.

\item  We generated 1000 light curves scrambling the observed rates while keeeping 
unchanged the temporal distribution of the bins. For each of them we have produced the 
periodogram in the 0.5--10 d time interval (49335 trial periods). The highest value obtained among all these 
periodograms is 97.9. Therefore, the probability of random occurrence for the observed $\chi^2$
value ($\sim240$) is lower than  $2\times 10^{-8}$ that corresponds to a significance of
P$_0$ higher than 5.5 standard deviations.

\end{enumerate}

\begin{figure}
\begin{center}
\centerline{\includegraphics[width=6cm,angle=-90]{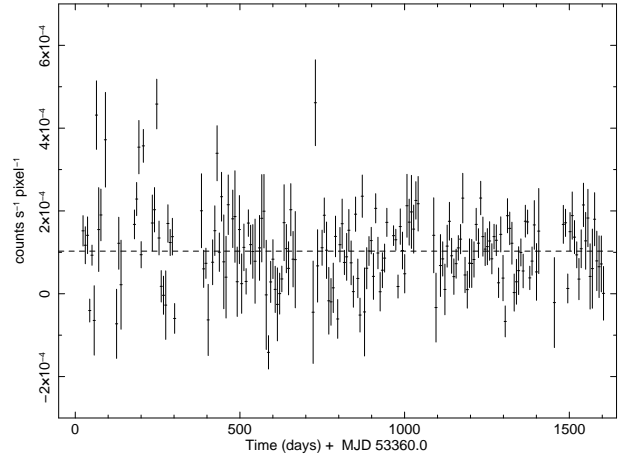}}
\caption[IGRJ16493-4348 BAT light curve]{15--50 keV BAT light
curve of IGRJ16493-4348 with a bin time P$_0=6.78$ d. The dashed line
represents  the average count rate.
                }
                \label{lc} 
        \end{center}
        \end{figure}

Figure~\ref{lc} shows the 15--50 keV light curve of \src, obtained excluding all
pointings where the source was at an off-axis angle higher than 40 deg, with a bin time of 
P$_0=6.78$ d, excluding bins with an exposure fraction less than 5\%.The average intensity in 
each P$_0$ bin shows a maximum variability of a factor $\sim4$. 

\subsection{XRT data}
The knowledge of the orbital period and of the middle eclipse time allows us to
determine that the \sw-XRT observation was performed at the orbital phase interval 0.45--0.51  
with respect to the epoch of the eclipse.
The 0.2--10 keV light curve shows a persistent emission with an average count rate of 
0.376$\pm$0.015 count s$^{-1}$ and a variability within a factor 3.

We applied the folding technique to search for short periodicities in the time 
range 5--1000 s.  The XRT SSB arrival
times were corrected for the pulsar binary motion using P$_0$, T$_{\rm eclipse}$, a semi-major axis 
$a\sim55\,R_{\sun} \sim 255,365$ lt-s (see Sect.~4 for its derivation), and 
assuming an orbit inclination of 90 degrees and an accentricity e=0.
 We found no evidence for any significant periodic signal.

\section{Discussion \label{discuss}}

We have presented the results obtained from the analysis of the data collected 
by \sw-BAT during the first 54 months of the Swift mission on 
the supergiant HMXB IGR~J16493--4348.
The source is detected at a significance level of $\sim$21 standard 
deviations.

The long monitoring coupled with a good observation duty cycle ($\sim13\%$ per day)
allowed us to unveil a periodicity of $\sim6.782\pm0.002$ d  that
we interpreted as the orbital period of the binary system.

%
The knowledge of the orbital period allows us to derive the the semi-major axis of the
binary system through the Kepler's third law: $a^3=G$ P$_0^2$$(M_{\star}+M_{\rm X})/4\pi^2$ 
where $M_{\star}$ and $M_{\rm X}$ are the masses of the supergiant and
compact object, respectively. Adopting $M_{\star}=47\,M_{\sun}$
\citep[][for an B0.5 Ib star, of radius $R_{\star}=32.2\,R_{\sun}$]{Searle2008}
and a typical neutron star mass of $M_{\rm X}=1.4\,M_{\sun}$
we obtain $a\sim55\,R_{\sun} \sim 2\,R_{\star}$.
This orbital separation is  common among classical sgHMXB \citep{Walter2007}.
The folded light curve shows the presence of a full eclipse with an estimated 
time of mid-eclispe at T$_{\rm eclipse}=(54175.92 \pm 0.10) \pm  n \times$ P$_0$ MJD and it appears flat outside the eclipse.
The duration of the eclipse ($\sim 12\%$ of the orbital period) is roughly consistent with the occultation
of the compact source by the companion star in a highly inclined orbit with low eccentricity.

\begin{figure}
\begin{center}
\centerline{\includegraphics[width=6cm,angle=-90]{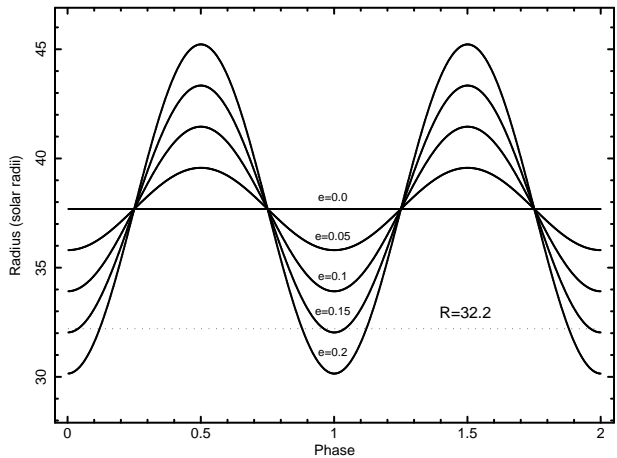}}
\caption[]{The Lagrangian point of the binary system as a function of the orbital
phase for different orbital eccentricity. The straight dashed line represents
the radius of the companion star. 
                }
                \label{lcc} 
        \end{center}
        \end{figure}

The 15--50 keV light curve of IGR~J16493--4348 is variable up to a factor $\sim$20 on timescales 
of $\sim$300 s while its average over an orbital period shows a variability 
up to a factor $\sim$4.
The inferred orbital separation suggests a wind-fed system.
This hypothesis and the radius of the supergiant allows 
to limit the eccentricity of the
orbit.  Figure~\ref{lcc}  shows how the Lagrangian point  between the two stars changes 
 as a function of the orbital
phase for different orbit eccentricities \citep{paczynski71}; the maximum eccentricity compatible 
with a wind-fed system (Lagrangian
point lower than the supergiant radius) is 0.15.

\section*{Acknowledgments}

This work was supported by contracts ASI I/088/06/0 and I/023/05/0.


\bsp

\label{lastpage}

\end{document}